# Synthesis of achiral rod-shaped triazolic molecules and investigation of their striped texture and propeller-patterned nematic droplets

Souria Benallou[a], Salima Saïdi-Besbes[a]*, Abdelatif Bouyacoub[b], Eric Grelet[c]*

[a] Université Oran 1 Ahmed Ben Bella, Laboratoire de Synthèse Organique Appliquée, Département de chimie, Faculté des sciences exactes et appliquées, BP 1524 EL Mnaouer, Oran, Algeria
E-mail: saidi.salima@univ-oran1.dz

[b] Université Oran 1 Ahmed Ben Bella, Laboratoire de Chimie Fine, Département de chimie, Faculté des sciences exactes et appliquées, BP 1524 EL Mnaouer, Oran, Algeria

[c] Centre de Recherche Paul-Pascal, UMR 5031, CNRS & Université de Bordeaux, 115 Avenue Schweitzer, F-33600 Pessac, France
E-mail: eric.grelet@crpp.cnrs.fr
ORCID: https://orcid.org/0000-0002-9645-7077

**Abstract**

In this study, novel achiral mesogens containing 4,4'-biphenyl central core connected on both sides through an ester function to 1-(4-(alkyloxy)phenyl)-1$H$-[1,2,3]-triazolyl group have been synthesized and characterized. Different mesophases have been identified, with the appearance of both a periodic striped pattern and propeller-patterned droplets close to the transition between the nematic and the smectic-C phases. This periodic texture has been found to be independent of the geometry of the sample cell. These observations suggest a spontaneous symmetry breaking which can result either from an intrinsic enantioselective self-sorting process with a chirality synchronization mechanism, or from an extrinsic surface mechanism at the interface between the two phases.

1. Introduction

Liquid crystals (LCs) are ubiquitous in our daily lives, found in a wide range of technological applications including display, optics, light-emitting diodes, photovoltaics, nanophotonics and biosensors [1-2]. These outstanding states of matter, which combine orientational and positional orders, continue to arouse the interest of the scientific community due to their potential for strategic applications. Numerous self-assembling systems built-up from functional organic molecules have been synthetized and investigated for their LC properties. The nature and stability of the resulting mesophases are closely dependent on the extent of intermolecular interactions and can be tuned and controlled through a suitable design of the molecular architecture and shape.

Rod-like (calamitic) mesogens have contributed extensively to LC development [3]. They are generally composed of a central rigid core containing at least two aromatic rings and one or more flexible alkyl chains and/or polar groups in terminal positions. Flexible connecting groups and lateral substituents can also be introduced to achieve the required criteria for liquid crystalline behaviour. Such anisotropic molecules usually give rise to the formation of typical nematic and smectic organizations.

The use of heterocyclic rings as a core unit in the molecular structure of LC materials is an efficient and convenient way to modify the physical, optical and electronic properties of LCs [4-5]. The presence of electronegative heteroatoms such as oxygen, nitrogen or sulphur can impart lateral and/or longitudinal dipoles to the designed LC compounds and affect substantially their geometry and shape, which in turn can modify their mesomorphic behaviour. N-heterocyclic compounds also have the ability to induce specific interactions such as hydrogen bonding, donor–acceptor assemblies or metal complexation favouring the formation of supramolecular mesogens [6-8].

[1,2,3]-triazole is a key building block in materials science for electronic, biomedical and supramolecular applications [9-10]. This five-membered N-heterocycle presents remarkable features in terms of chemical and thermal stability, polarity and the ability to interact with different biological, organic and inorganic targets [11-12]. In recent years, a growing number of thermotropic triazole based liquid crystalline materials have been designed, synthesized and characterized to highlight the structure–property correlation [13-16]. Several studies emphasized the role of the linking groups in influencing the molecular architecture and polarity of LC materials and consequently their mesomorphic properties. For instance, ester and alkene groups have been shown to promote more layer organization than the oxymethylene group. This arises from the lack of linearity of the related mesogen, which disrupts the lamellar ordering [17-18]. It is also generally noticed that liquid crystalline organizations and mesophase thermal

stability are improved by the insertion of a polar connector due to the enhancement of the dipole moment in the mesogen [19].

In this study, we develop a new molecular design approach to obtain mesogenic molecules composed of a 4,4'-biphenyl central core connected on both sides through an ester function to a 1-(4-(alkyloxy)phenyl)-1*H*-[1,2,3]-triazolyl group with varying number carbon atoms in the terminal alkyl chains n = 8-12 (**1a-c**) as shown in Scheme 1. The mesomorphic properties of these compounds are investigated by means of X-ray scattering, optical microscopy and calorimetry. These mesogens, which do not exhibit any asymmetric tetrahedral carbons (i.e. do not have any chiral center), show a periodic striped pattern phase between the nematic and smectic-C states, as well as unusual propeller-patterned droplets. The mechanisms at the origin of these unusual features will be discussed.

## 2. Experimental section

### 2.1. Synthesis and chemical characterization

All the reagents have been purchased from Aldrich and used as received. The solvents are of commercial grade quality and have been dried and distilled before use. 1-(4-Alkyloxyphenyl)-1-H-[1,2,3]-triazole-4-carboxylic acid have been prepared according to a recently described procedure [20]. $^1$H and $^{13}$C NMR spectra have been recorded on a Bruker 300 MHz spectrometer (Wissembourg, France). Tetramethylsilane has been used as an internal reference for chemical shifts. Column chromatographies have been carried out using Merck silica gel (Kieselgel 60, 230–400 mesh) as the stationary phase (Merck Millipore, Gernsheim, Germany). High resolution mass spectroscopy (HRMS) has been carried out using a Bruker Daltonics - micrOTOF-Q mass spectrometer coupled to LC system in ESI mode.

**1a-c:** 2.5 mmol of 4-(alkyloxyphenyl)-1*H*-[1,2,3]-triazole-4-carboxylic acid and 25 mmol of thionyl chloride are stirred at 85°C under nitrogen atmosphere for 5-6 hours. Thereafter, the excess of thionyl chloride is evaporated and a solution of triethylamine (7.2mmol) in 10 mL of freshly distilled dichloromethane is added. The mixture is cooled using an ice bath and the reaction is stirred under nitrogen atmosphere until a homogeneous solution is obtained. 4,4'-biphenol (1.25mmol) dissolved in 5 mL of dry dichloromethane is then added dropwise and the mixture is stirred at 40°C for 48-72 hours. After cooling the mixture, water is added and the organic layer is extracted, washed three times with 1M HCl solution then with water until neutrality and finally dried under anhydrous MgSO$_4$. The solvent is evaporated under reduced pressure and the crude product is purified by column chromatography using as eluent dichloromethane/ethyl acetate (20/1).

*bis (1-(4-octyloxyphenyl)-1H-[1,2,3]-triazole)-4,4'-biphenyl dicarboxylates (1a)*

Yield (56%); $^1H$ *NMR (300 MHz, CDCl$_3$):* $\delta$ = 0.92 (t, 6H, CH$_3$, $^3J$= 6.71 Hz), 1.32 (m, 20H, CH$_3$(CH$_2$)$_5$), 1.85 (m, 4H, OCH$_2$CH$_2$), 4.07 (t, 4H, OCH$_2$, $^3J$ = 6.58 Hz), 7.06 (d, 4H, Ar-H, $^3J$ = 9.04 Hz), 7.39 (d, 4H, Ar-H, $^3J$ = 8.65 Hz), 7.72 (m, 8H, Ar-H), 8.61 (s, 2H, H$_5$-triazole); $^{13}C$ *NMR (75MHz, CDCl$_3$):* $\delta$ = 14.11, 22.67, 26.00, 29.34, 31.81, 68.58, 115.51, 122.00, 122.52, 128.34, 129.38, 137.41, 139.79, 149.78, 154.66, 159.12; HRMS (ESI) *m/z* calculated for C$_{46}$H$_{53}$N$_6$O$_6$ ([M+H]$^+$): 785.3985; found: 785.4021.

*bis (1-(4-decyloxyphenyl)-1H-[1,2,3]-triazole)-4,4'-biphenyl dicarboxylates (1b)*

Yield (48%); $^1H$ *NMR (300 MHz, CDCl$_3$):* $\delta$ = 0.91 (t, 6H, CH$_3$, $^3J$ = 6.71 Hz), 1.30 (m, 28H, CH$_3$(CH$_2$)$_7$), 1.85 (m, 4H, OCH$_2$CH$_2$), 4.05 (t, 4H, OCH$_2$, $^3J$ = 6.58 Hz), 7.09 (d, 4H, Ar-H, $^3J$ = 9.04 Hz), 7.39 (d, 4H, Ar-H, $^3J$ = 8.65 Hz), 7.69 (m, 8H, Ar-H), 8.61 (s, 2H, H$_5$-triazole); $^{13}C$ *NMR (75MHz, CDCl$_3$):* $\delta$ = 14.13, 22.69, 25.99, 29.57, 31.91, 68.58, 115.51, 122.03, 122.50, 128.33, 129.38, 138.45, 139.78, 149.91, 159.12, 160.14; HRMS (ESI) *m/z* calculated for C$_{50}$H$_{61}$N$_6$O$_6$ ([M+H]$^+$): 841.4608; found: 841.4647.

*bis (1-(4-dodecyloxyphenyl)-1H-[1,2,3]-triazole) 4,4'-biphenyl dicarboxylates (1c)*

Yield (61%); $^1H$ *NMR (300 MHz, CDCl$_3$):* $\delta$ = 0.90 (t, 6H, CH$_3$, $^3J$ = 6.71 Hz), 1.29 (m, 36H, CH$_3$(CH$_2$)$_9$), 1.87 (m, 4H, OCH$_2$CH$_2$), 4.05 (t, 4H, OCH$_2$, $^3J$ = 6.58 Hz), 7.09 (d, 4H, Ar-H, $^3J$ = 9.04 Hz), 7.36 (d, 4H, Ar-H, $^3J$ = 8.65 Hz), 7.69 (m, 8H, Ar-H, $^3J$=9.17); 8.61 (s, 2H, H$_5$-triazole); $^{13}C$ *NMR (75MHz, CDCl$_3$):* $\delta$ = 14.13, 22.70, 26.00, 29.36, 31.92, 68.58, 115.51, 122.02, 122.50, 128.33, 129.37, 138.54, 139.77, 147.14, 159.12, 160.13; HRMS (ESI) m/z calculated for C$_{54}$H$_{69}$N$_6$O$_6$ ([M+H]$^+$: 897.5261; found: 897.5273; C$_{54}$H$_{68}$N$_6$NaO$_6$ ([M+Na]$^+$: 919.5080; found: 919.5093.

## 2.2. Methods for mesomorphic property determination

The melting points, transition temperatures and phase transition enthalpies have been determined using a differential scanning calorimetry (DSC Q200®, TA Instruments). The mesomorphic textures have been observed using BH-2 and BX-51 Olympus microscopes equipped with a Linkam THM600 heating stage and with a colored JAI CV-M7 and CoolSNAP ES (Photometrics) cameras, respectively.

X-ray diffraction experiments have been performed by using a rotating anode generator (Rigaku Nanoviewer MicroMax 007HF) coupled to a confocal Max-Flux® Osmic mirror (Applied Rigaku Technologies, Austin, USA) producing a beam with a wavelength of 1.54 Å together with a homemade heating stage with a thermal stability of 0.1°C. The spectra have been recorded with a bi-dimensional detector (MARResearch-345). All samples were introduced as a powder in glass capillary tubes (Glas, Muller, Germany) having a diameter of 1.5mm. The spectra were analyzed using FIT2D software (ESRF; http://www.esrf.eu/).

3. Results

The targeted compounds have been synthetized according to the synthetic strategy outlined in Scheme 1. They were obtained from 1*H*-[1,2,3]-triazole-4-carboxylic acid intermediates, reported in a previous published work [20], which have been transformed into the corresponding triazoyl chloride compounds. The condensation with 4,4'-biphenol affords the final compounds with good yields. The purity and more specifically the absence of chiral impurities has been checked by NMR spectroscopy and HRMS (See Experimental section).

The phase behavior and mesomorphic properties of the series **1a-c** have been determined by the combination of different experimental techniques, from X-ray diffraction (Figure 1) and optical microscopy (Figure 2) to calorimetry (Figures S1, S2, S3 and Table 1).

Beyond a set of crystalline states at low temperature, all compounds **1a-c** exhibit a smectic phase above which appears a nematic mesophase. The physical signature of the smectic phase has been obtained by X-ray diffraction, where the X-ray scattering spectra show a first sharp Bragg peaks in the range of $q_{100}$ = 1.95 −2.00 nm$^{−1}$ and the second-order reflection (200) confirming the lamellar organization (Figure 1). The signal of the disordered aliphatic chains corresponds to the broad reflection at about 13.6 nm$^{−1}$. The increase of the smectic layer spacing d = $2\pi/q_{100}$ with temperature (Inset of Figure 1) stems from the rise of the tilt angle of the molecular director due to thermal fluctuations, and therefore suggests a smectic-C like order. The lamellar spacing value of about d = 3.2 nm is about the same size as the rigid core of the mesogen (See Scheme 2 for the theoretical calculation considering the energy minimized structure obtained by DFT calculations). Above the smectic range, the recorded scattered X-ray spectra do not show sharp Bragg reflections but exhibit a very broad peak at low range of scattering wavevectors q, which is the hallmark of molecular liquid-like order consistent with the reported nematic and isotropic liquid phases at high temperature.

Polarized-light optical microscopy analysis has been carried out to complete the identification of the different mesophases. The micrographs for compound **1b** sandwiched between a glass slide and a cover slip are shown in Figure 2. Upon cooling down from the isotropic liquid, characteristic nematic swarming is observed that evolves to birefringent nematic texture with Schlieren defects (Fig. 2(a)). At lower temperature, a striped pattern texture appears as depicted in Figures 2(c) below which the fan shape texture or Schlieren pattern characteristic of a tilted smectic phase is seen (Fig. 2(e)).

The presence of transient state between nematic and smectic mesophases, i.e. in the temperature range where the striped patterns are observed, can be inferred from some but not all differential scanning calorimetry thermograms (Figs. S1 to S3). Indeed, a small endothermic peak can be barely guessed from a rugged line with structureless broad features only in the first heating cycles at 225 and 228 °C, for compounds **1a** (Figure S1) and **1b** (Figure S2), respectively. However, this peak is more difficult to detect during the second cycle, which seems to indicate some thermal damage to the sample (Figs. S1, S2, S3).

Despite its lack of clear thermodynamic signatures, the striped textures have been observed enantiotropic ally, i.e. on both heating and cooling and appear to be independent of the nature of the substrates used (glass, silicon wafer, rubbed Teflon coating), and the sample geometry (glass cell or open droplets, thin or thick samples from a few to tens of µm). Note that these periodic structures have been found over many heating-cooling cycles. Furthermore, the striped pattern stands out for its high fluidity (see supplementary movie) particularly on heating. Similar behavior has been seen for each compound of the investigated series. The repeat distance between parallel bright and dark lines corresponding to the periodicity of the stripes does not show significant variation within observation times of a few minutes, regardless of the studied domain. The stripe period p is shown to be temperature dependent and increases with decreasing temperature, as reported in Figure 3, in agreement with the behavior commonly encountered for the chiral nematic (cholesteric) mesophase. However, the usual divergence of the cholesteric periodicity scaling with $p \propto (T - T_c)^{-\nu}$ with $\nu > 0$ and $T_c$, an effective critical temperature related to the phase transition temperature, is not observed here when approaching the smectic phase transition [21]. Instead of a convex diverging function, a concave one fits our data with $p \propto (T_c - T)^\nu$ (Eq. (1)) and $\nu \approx 0.6$, demonstrating an anomalous behavior for the stripe periodicity.

To further investigate this striped pattern, a very thin supported droplet geometry has been employed for compound **1a** (Figure 4). Upon cooling from the nematic phase, right-handed and left-handed propeller patterns appear in these droplets, as illustrated in Figure 4. They are observed in a temperature range from 207 to 220°C, which is within the striped texture range obtained during very slow cooling processes, where supercooling shifts all transition to lower temperatures by a few °C. It is worth mentioning here that no coexistence of adjacent domains with opposite handedness has been detected by optical microscopy.

## 4. Discussion

A striped texture is a common feature observed at the N-SmC phase transition, as well as at the Iso-SmC transition. These so-called transition bars were discovered in the 1970s and attributed to spontaneous undulations at the SmC-N interface [22-25]. The stripes stem from the coupling between the surface tension anisotropy and the director orientation in the SmC layers [22-29]. In such a mechanism, the undulation repeat distance depends on the sample thickness, the thermal gradient and the presence of impurities [23,26]. The less pure the product is, the broader the two-phase coexistence region is and the more pronounced the stripes are. Even if the purity of the compounds has been carefully checked by NMR and HRMS, the formation of impurity traces at high temperature for which the striped patterns are observed cannot be formally excluded. Such impurities may be generated, especially during prolonged and repeated heating steps. Moreover, the high temperatures of the N-SmC phase coexistence favor the presence of (vertical) thermal gradients from the heating stage. Combined with both the lack of a clear transition peak in the DSC thermograms and the unusual stripe periodicity dependence (Figure 3), this suggests that the striped patterns of **1a-c** mesogens could have an extrinsic origin, based on surface induced symmetry breaking.

Nevertheless, the unambiguously observed propeller-patterned droplets as shown in Figure 4 also suggests another mechanism based on an intrinsic mirror symmetry breaking of the achiral mesogens to form chiral nematic domains. Furthermore, the striped texture could then be interpreted as cholesteric fingerprints, providing a measurement of the helical pitch of the sample. In recent decades, the phenomenon of mirror-symmetry breaking has been reported in several liquid crystalline phases and particularly in layered phases [30-40]. Even if the archetypal mesogens inducing such enantioselective segregation are achiral bent-core molecules and bent-shape mesogenic dimers that can acquire conformational chirality in layered smectic mesophases [41-43], propeller-like molecular conformation has also been found with achiral molecules through specific non covalent interactions like hydrogen bonding [44], coordinative interactions and π-π interactions between aromatic cores [44-46].

The existence of propeller-pattern droplets in the series of achiral compounds **1a-c** with only homochiral domains may indicate the spontaneous formation and resolution of enantiomeric species from the nematic phase with a dynamic process of chirality synchronization. That would imply a stochastic generation of an excess of one enantiomer in the liquid crystalline state and a subsequent chirality amplification to a given homochiral domain or droplet, thanks to collaborative effect that stabilize the molecules with identical chiral conformation.

The unusual formation of propeller-patterned droplets for the series reported here is unexpected since these mesogens are achiral molecules having neither a chiral center nor any functional group likely to induce self-assembly through hydrogen bonding interactions. Interestingly, analogous compounds described in a previous work bearing oxymethylene connector function instead of the ester one, reveal neither striped structures nor propeller-patterned droplets (only the enantiotropic smectic-C mesophase is seen above 220°C without any nematic state) [18]. This result may suggest the key-contribution of the ester group on the induction of such features. Previous reports have shown peculiar behavior of ester molecules in chirality amplification [47]. Takezoe *et al*. proved that rod-shape molecules bearing an ester function can act as a chiral dopant in a cholesteric guest phase and induce more than a 10% increase of the twisting power [48-49]. Enantioselective segregation of achiral nematic phase into chiral domains of opposite handedness has been also reported for achiral bent-core systems based on bis(phenyl)oxadiazole compounds with two ester linkages [39]. For our investigated compounds, a collective effect implying the specific role of ester groups in the intermolecular interactions could be invoked for the stabilization of one of the axial chiral conformers in series **1a-c** (Scheme 2).

5. **Conclusion**

In this work, the synthesis and the mesomorphic characterization of achiral bis-triazole substituted biphenyl derivatives have been performed, from which two liquid crystalline phases, i.e. nematic and smectic-C mesophases, have been identified. At the transition between the two mesophases, a spontaneous striped texture is shown. This periodic structure could be associated with a spontaneous symmetry breaking, having either an extrinsic origin resulting from the instability of the interface at the phase coexistence, or an intrinsic one, promoting the stochastic emergence of one chiral conformer as well as the development of macroscopic homochiral nematic domains. The latter interpretation is supported by the formation of thin propeller-like patterned droplets, whose helicity can be inverted through thermal annealing.

Further experimental studies are required to fully explore the mechanism of this spontaneous symmetry breaking evidenced in this series, implying the design of compounds with lower transition temperatures.

**Author contributions**

SB and SSB designed and synthesized the compounds, AB performed the molecular simulations, EG and SSB performed the physical characterizations, interpreted the data and wrote the manuscript.

The authors report there are no competing interests to declare.

**Acknowledgements**

This work was supported by DGRSDT and CNRS.

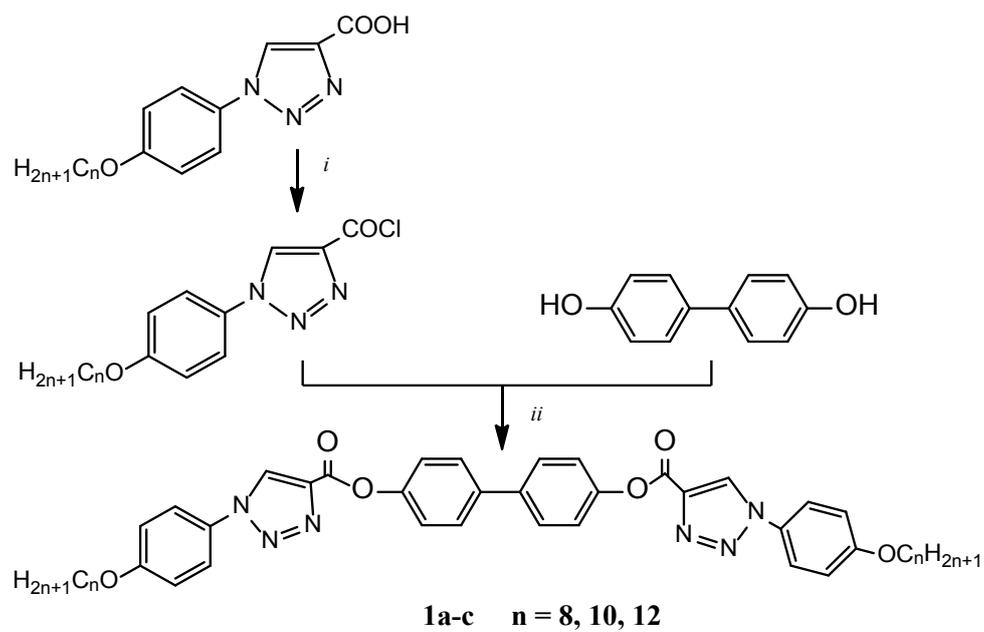

***Scheme 1.*** *Conditions for the synthesis of triazole derivatives (**1a-c**) (i) SO₂Cl₂, 85°C, 5-6 h (vi) N(C₂H₅)₃, DCM, 40°C, 48-72 h.*

| Compound | n | Transitions temperatures (°C) [transition enthalpies ΔH (kJ.mol$^{-1}$)] |
| --- | --- | --- |
| **1a** | 8 | Cr(193)[32.6] SmC(225)[2.5] X(252)[0.49] N(283)[0.5] I |
| **1b** | 10 | Cr(147)[7.8] SmC(228)[3.3] X(242)[1.57] N(272)[0.3] I |
| **1c** | 12 | Cr (168)[42.11] SmC(258)[4.8] X(281)[5.1] N+I |

***Table 1.*** *Transition temperatures (°C) and related enthalpy values ΔH (kJ.mol$^{−1}$) of compounds **1a-c** determined by DSC (10°C/min) during the first cycle. X stands for the striped pattern phase.*

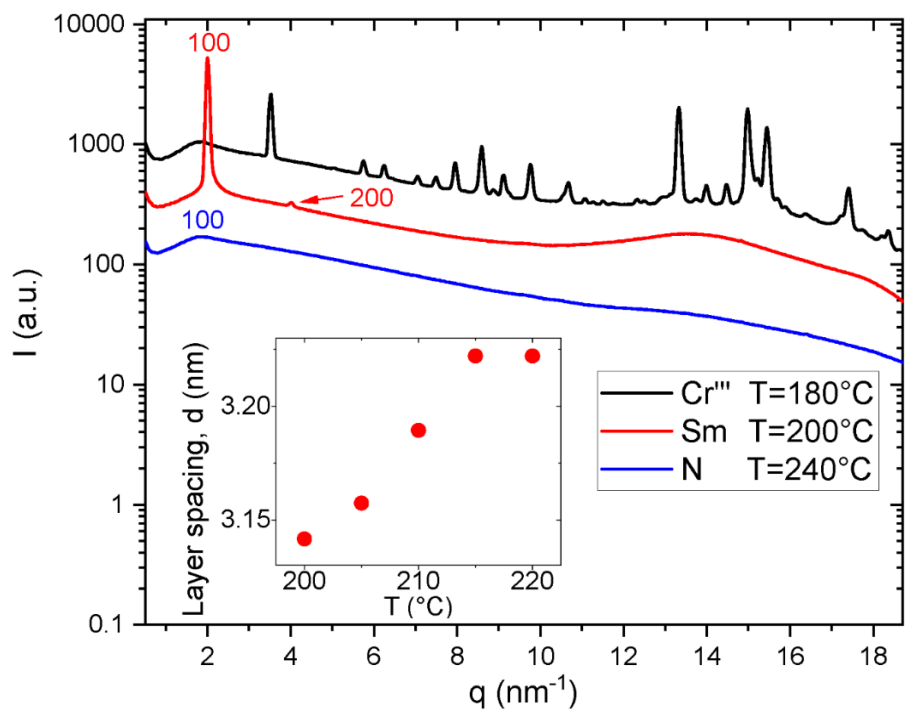

*Figure 1.* X-ray diffractograms showing the main phases found on the compound *1a* at different temperatures. The crystalline phase displays many sharp reflections, as indicated by the black line. The smectic mesophase (red line) has two main Bragg reflections related with the layer periodicity ($d=2\pi/q_{100}$) and a broad peak stemming from the disordered aliphatic chains at about 14 nm$^{-1}$. At higher temperature, the spectrum (blue line), typical of the liquid-like order characteristic of a nematic phase is shown. Inset: Increase of the smectic layer spacing d with temperature indicating tilted molecules in the layers i.e. a smectic-C like local order.

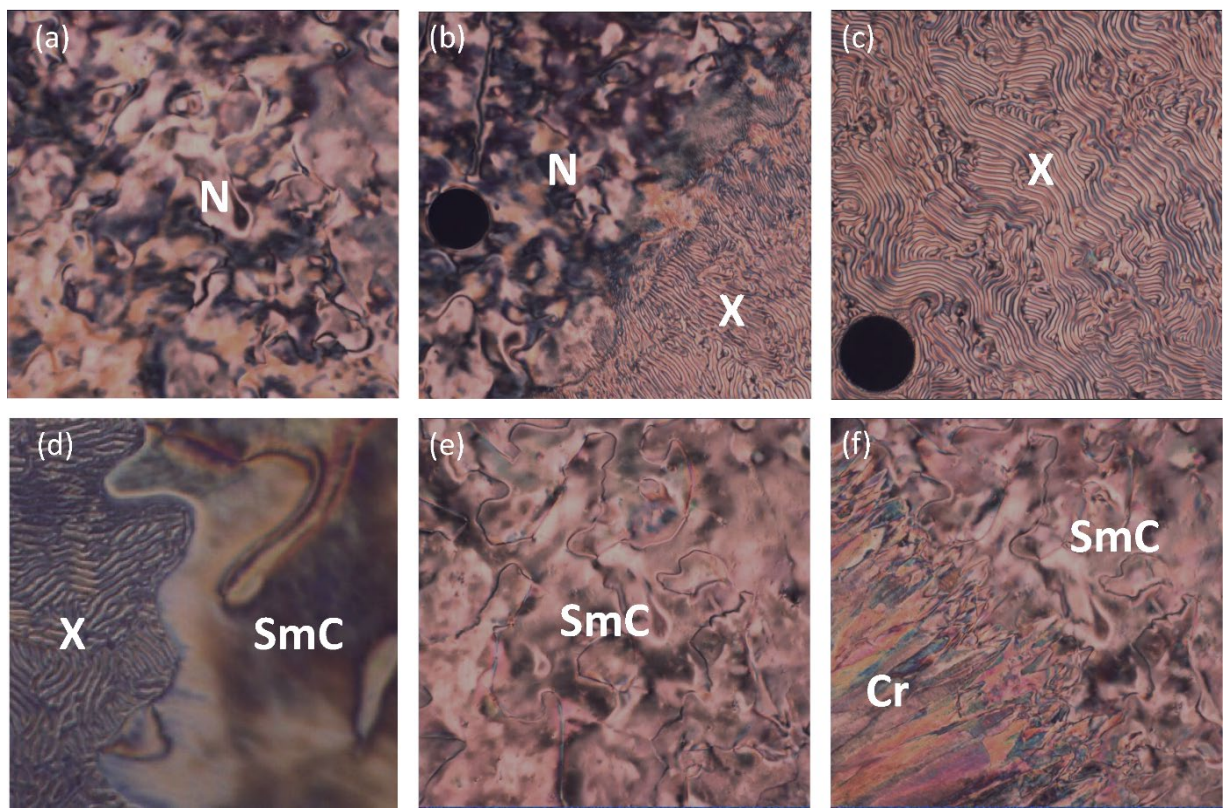

***Figure 2.*** *Optical textures observed between crossed polarizers on cooling at 5°C/min with the compound **1b** prepared in a glass cell. Both nematic N (a) and smectic-C (e) phases exhibit a Schlieren texture, whereas the X phase (c) displays a striped texture from which the period can be measured. Note the phase coexistence between all the different phases; (b): N-X; (d): X-SmC; (f): SmC-Cr. Image size: 330 X 330 μm².*

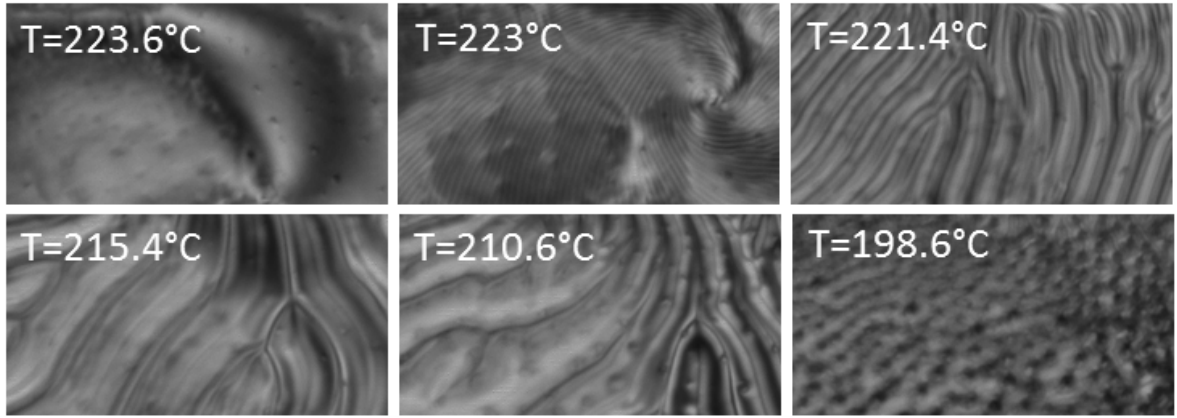

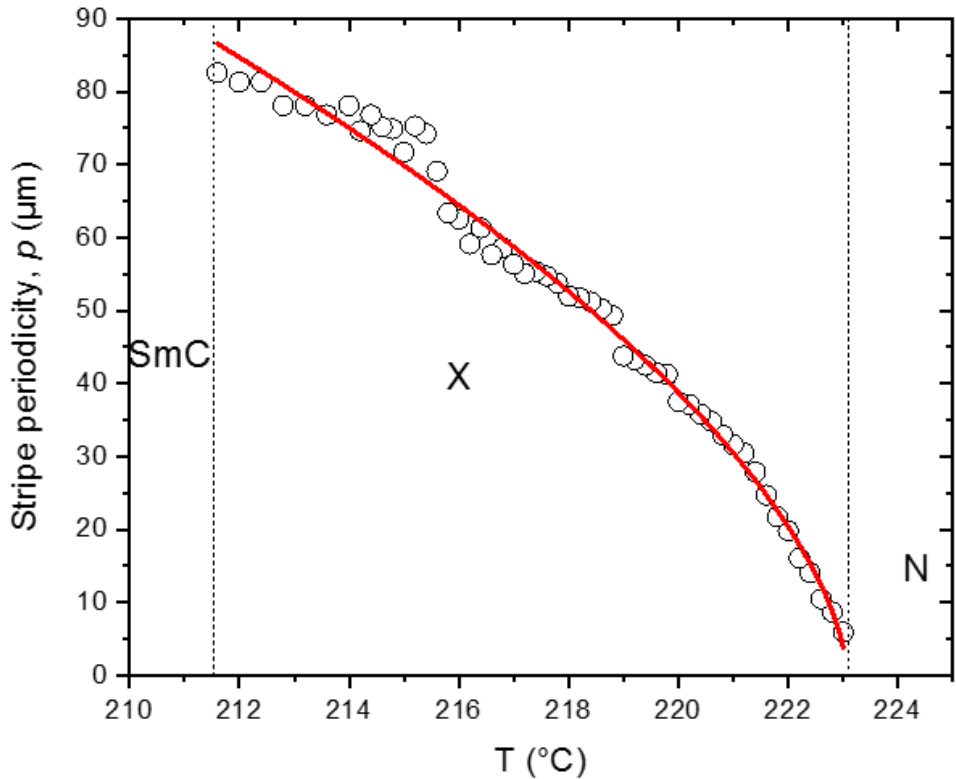

*Figure 3. (Up) Pictures of liquid crystalline phase sequence observed between crossed polarizers on cooling at 0.2°C/min the sample **1a** in the geometry of open droplet. The sample starts from a nematic phase (223.6°C) with a typical Schlieren texture to a smectic-C phase (198.6°C) via the striped pattern phase (X), whose periodicity increases with decreasing temperature. Image size: 150 X 90 μm². (Bottom) Temperature dependence of the undulation period measured on cooling sample **1a** (black circles). The gaps in the measurements stem from microscope focus tuning of the observation region when temperature is changed during the experiment. The fit to Eq. (1) is shown by the solid red line.*

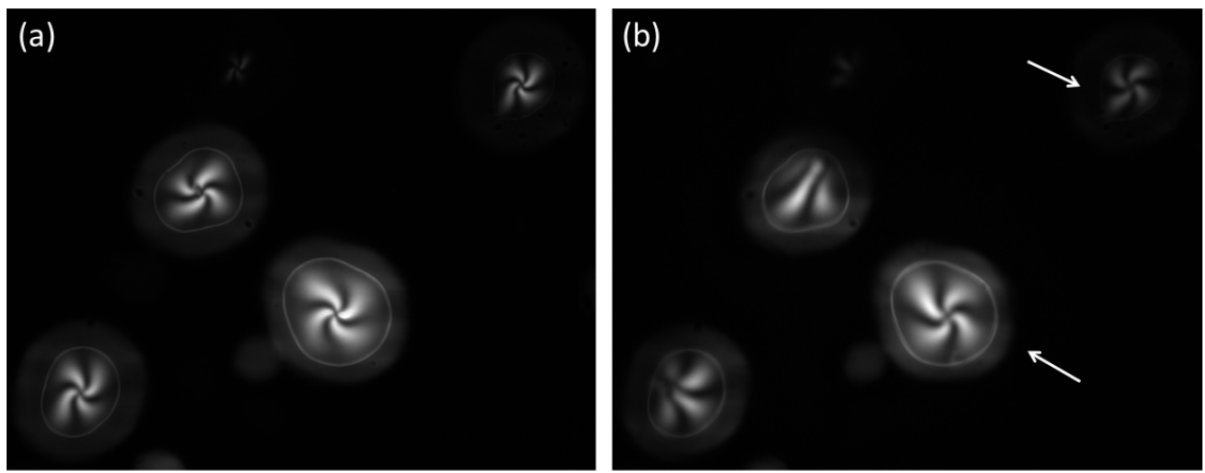

*Figure 4.* Propeller-patterned droplets observed for compound *1a* between crossed polarizers by optical microscopy. (a) Mixture of left and right-handed droplets obtained in a few µm thick sample on slow cooling at T=210°C. (b) Same droplets after a thermal annealing in the isotropic liquid phase and cooled down back at 210°C. The handedness of the propeller-pattern has been inverted for two droplets (indicated by white arrows) while the others do not exhibit chiral inversion. Image size: 280 X 210µm².

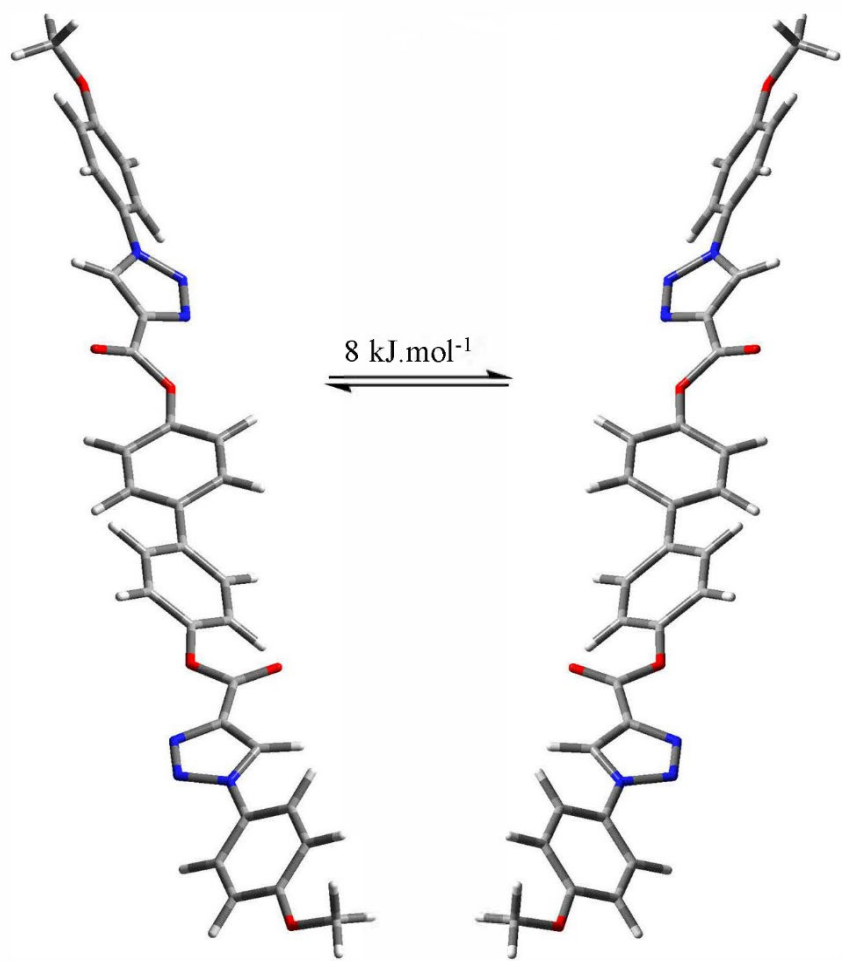

***Scheme 2.*** *Energy minimized structure of compound **1a-c** for which the aliphatic chains have not been represented, and its mirror image conformer according to density functional theory (DFT) calculations. The activation energy for enantiomerization of the helical conformers is 8 kJ.mol$^{-1}$. Note that this energy barrier is rather weak, and is not* a priori *expected to prevent the chiral conformational interconversion.*